\documentclass[twocolumn,showpacs,preprintnumbers,amsmath,amssymb,aps]{revtex4}
\usepackage{graphicx}% Include figure files
\usepackage{dcolumn}% Align table columns on decimal point
\usepackage{bm}% bold math

\begin{document}

%\preprint{APS/123-QED}

\author{Peter B. Weichman}

\affiliation{$^2$ALPHATECH, Inc., 6 New England Executive Place,
Burlington, MA 01803}

\title{Surface modes and multi-power law structure in the early-time
electromagnetic response of magnetic targets}

\date{\today}

\begin{abstract}

It was recently demonstrated [P. B. Weichman, Phys.\ Rev.\ Lett.\
{\bf 91}, 143908 (2003)] that the scattered electric field from
highly conducting targets following a rapidly terminated
electromagnetic pulse displays a universal $t^{-1/2}$ power law
divergence at early time.  It is now shown that for strongly
permeable targets, $\mu_c/\mu_b \gg 1$, where $\mu_b$ is the
background magnetic permeability, the early time regime separates
into two distinct power law regimes, with the early-early time
$t^{-1/2}$ behavior crossing over to $t^{-3/2}$ at late-early
time, reflecting a spectrum of magnetic surface modes. The latter
is confirmed by data from ferrous targets where $\mu_c/\mu_b =
{\cal O}(10^2)$, and for which the early-early time regime is
invisibly narrow.

\end{abstract}

%\keywords{Suggested keywords}% Use showkeys class option if keyword
                              % display desired
\maketitle

Remote characterization of buried targets is a key goal in many
environmental geophysical applications, such as landmine and
unexploded ordnance (UXO) remediation \cite{serdp}. A tool of
choice is the time-domain electromagnetic (TDEM) method, in which
an inductive coil transmits low frequency [typically ${\cal
O}(100\ {\rm Hz})$] EM pulses into the ground. Following each
pulse [terminated rapidly on a ramp timescale $\tau_r = {\cal
O}(10^2 \mu{\rm s})$], the voltage $V(t)$ induced by the scattered
field is detected by a receiver coil. Standard TDEM sensors are
capable of resolving signals from very small (of order 1 gram)
metal targets \cite{NH01}, and are therefore well suited to UXO
detection.

Low frequency yields increased sensitivity to conducting targets,
and increased exploration depth (below 5 m) but leads to nearly
complete loss in spatial resolution. Lacking direct spatial
imaging capability (enabling straightforward identification), one
is reduced to seeking such information indirectly via a careful
analysis of the full time dependence of $V(t)$. As described in
Refs.\ \cite{WL03,W03}, this signal is affected by both
\emph{intrinsic} (target size, shape, geometry, and other physical
characteristics) and \emph{extrinsic} (relative target-sensor
position and orientation, transmitter and receiver coil
geometries, pulse waveform, etc.)\ properties, and the key to
discrimination is the extraction of the former from the
``background'' of the latter.  The aim of this letter is to
further develop such formalism for the early time part of $V(t)$
\cite{W03}.

Conductor electrodynamics are essentially diffusive, and the basic
time scale $\tau_c = L_c^2/D_c$ is determined by the target
diameter $L_c$, and diffusion constant $D_c = c^2/4\pi\mu_c
\sigma_c$ (in Gaussian units) depending on the (relative) target
permeability $\mu_c$ and conductivity $\sigma_c$. Ferrous targets
(e.g., steel) are typically modelled as paramagnetic targets with
very large permeability $\mu_c = {\cal O}(10^2)$, and (MKS)
conductivity $\sigma_c = {\cal O}(10^7\, {\rm S/m})$ . Thus, even
for targets as small as $L_c = 1$ cm, one finds decay times
$\tau_c = {\cal O}(10^2\, {\rm ms})$, much larger than typical
pulse periods. Larger UXO-like targets, which are usually ferrous,
have even larger $\tau_c$. It will often be the case, therefore,
that the \emph{full measured range} of $V(t)$ will lie in the
early time regime, $t \ll \tau_c$. On the other hand, for a
nonmagnetic (e.g., aluminum) target of similar size one obtains
$\tau_c = {\cal O}(1\ {\rm ms})$, and the measured signal will
cover a much broader dynamical range.

\begin{figure}[tbh]

%use the following for two-column format
\includegraphics[scale=0.3,angle=270]{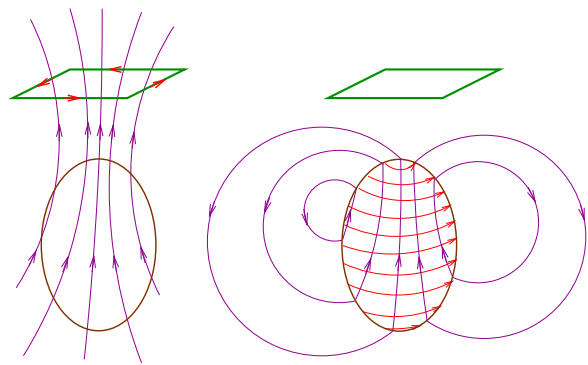}

%use the following for preprint format
%\includegraphics[scale=0.6,angle=270]{prlfig1.eps}

\caption{(COLOR) Schematic diagram of measurement and early time
dynamics. Left:  prior to pulse termination, transmitter coil
currents generate a magnetic field in the neighborhood of the
target. Right: just after pulse termination, the target interior
has not had time to adjust to the absence of the transmitted
field, and screening surface currents are generated to enforce the
correct EM boundary conditions.}

\label{fig1}
\end{figure}

\begin{figure}[tbh]

%use the following for two-column format
\includegraphics[scale=0.4]{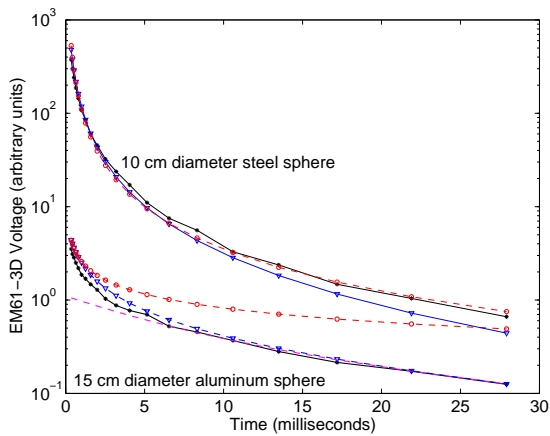}

%use the following for preprint format
%\includegraphics[scale=0.75]{prlfig2.eps}

\caption{(COLOR) Geonics EM61-3D data (black curves) from
nonmagnetic (aluminum) and ferrous (steel) targets, together with
predicted exact solutions \protect\cite{Jackson} (blue curves).
Asymptotic early-time power laws (red curves) are seen, with
$t^{-1/2}$ for aluminum, $t^{-3/2}$ for steel. The former shows a
crossover to multi-exponential decay behavior (magenta curve---the
first 60 slowest decaying modes), consistent with $\tau_e = {\cal
O}(10^{-2}{\rm s})$, $\tau_c = {\cal O}(10^{-1}{\rm s})$. The
latter covers essentially the entire observed range, consistent
with $\tau_e = {\cal O}(10^{-5}{\rm s})$, $\tau_c = {\cal O}(1\
{\rm s})$.}

\label{fig2}
\end{figure}

The early time response is governed by the dynamics of the
screening currents, induced in response to the rapid quenching of
the transmitted magnetic field immediately following pulse
termination (see Fig.\ \ref{fig1}). The initial diffusion of these
currents inward from the target surface generates a $t^{-1/2}$
power law in $V(t)$, with coefficient reflecting the surface
properties of the target \cite{W03}. However, underlying this
power law is the assumption that $t/\tau_c$ is the smallest
parameter in the problem.  It will now be shown that the
background-target permeability contrast $\mu_b/\mu_c$ can generate
a new small parameter that greatly limits the its range of
validity. Specifically, an extended calculation is described that
divides the early time regime $t < \tau_e$, where $\tau_e \ll
\tau_c$ is the point at which bulk effects first begin to enter
(see further below), into an early-early time regime, $0 < t \ll
\tau_{\rm mag}$, where the $V(t) \sim t^{-1/2}$ remains valid, and
a late-early time regime $\tau_{\rm mag} \ll t \ll \tau_e$, where
a new  power law $V(t) \sim  t^{-3/2}$ obtains. In the
neighborhood of the \emph{magnetic crossover time},
\begin{equation}
\tau_{\rm mag} = \tau_c (\mu_b/\mu_c)^2,
\label{1}
\end{equation}
an interpolation between the two power laws occurs (that is
visible only if $\tau_{\rm mag} \ll \tau_e$), for which the full
functional form is provided. For ferrous targets $\tau_{\rm
mag}/\tau_e = {\cal O}(10^{-4})$, and the early-early time regime
is essentially invisible, and only the $t^{-3/2}$ behavior is
seen, consistent with measured data: see Fig.\ \ref{fig2}.

At low frequencies the dielectric function in the ground and in
the target is dominated by the its imaginary part, $\epsilon =
4\pi i \sigma/\omega$, where $\sigma({\bf x})$ is the dc
conductivity, and the Maxwell equations may be reduced to a single
equation for the vector potential,
\begin{equation}
\nabla \times \left(\frac{1}{\mu} \nabla \times {\bf A} \right)
+ \frac{4\pi \sigma}{c^2} \partial_t {\bf A}
= \frac{4\pi}{c} {\bf j}_S,
\label{2}
\end{equation}
with magnetic induction ${\bf B} = \nabla \times {\bf A}$, field
${\bf H} = {\bf B}/\mu$, and gauge chosen so that the electric
field is ${\bf E} = -(1/c) \partial_t {\bf A}$.  The transmitter
loop generates the source current density ${\bf j}_S({\bf x},t)$.
The conductivity and permeability are separated into background
[$\sigma_b({\bf x})$, $\mu_b({\bf x})$] and conducting target
[$\sigma_c({\bf x})$, $\mu_c({\bf x})$] components, where
$\sigma_c$, $\mu_c$ vanish outside the target volume $V_c$. High
conductivity contrast, $\sigma_b/\sigma_c \ll 1$, is required, but
$\mu_b/\mu_c$ is arbitrary.

Equation (\ref{2}) is a vector diffusion equation, with contrast
$D_b/D_c = {\cal O}(10^7\mbox{--}10^9)$.  The ``background
communication time'' between instrument and target separated by
distance $R$ is $\tau_b = R^2/D_b$. For representative values
$\sigma_b = 0.1$ S/m, $\mu_b = 1$, $R = 3$ m, one obtains $\tau_b
= {\cal O}(10^{-6}{\rm s})$, instantaneous even on the scale of
$\tau_r$.  Thus, the electrodynamics of the background may be
treated as \emph{quasistatic}:  outside the target one may drop
the $\partial_t {\bf A}$ term in (\ref{2}). Further, following
pulse termination ($t > 0$) one has ${\bf j}_S \equiv 0$, hence
$\nabla \times {\bf H} = 0$, and one may derive ${\bf H} = -\nabla
\Phi$ from a scalar potential $\Phi$, which in turn satisfies
$\nabla \cdot (\mu_b \nabla \Phi) = 0$. The external field is
therefore entirely determined by the internal field through the
boundary condition \cite{Jackson} ${\bf \hat n} \cdot [\mu_b {\bf
H}_b - \mu_c {\bf H}_c] = 0$, where ${\bf H}_b$, ${\bf H}_c$ are
the fields just outside and inside the target surface, and ${\bf
\hat n}$ is the local surface normal.  Thus, one obtains a
Neumann-type boundary condition, $-{\bf \hat n} \cdot \nabla \Phi
= (\mu_c/\mu_b) {\bf \hat n} \cdot {\bf H}_c$, and hence the
formal solution
\begin{equation}
\Phi({\bf x}) = \int_{\partial V_c} d^2r g_N({\bf x},{\bf r})
\frac{\mu_c}{\mu_b} {\bf \hat n} \cdot {\bf H}_c({\bf r}),
\label{3}
\end{equation}
where $\partial V_c$ is the target surface, and $g_N$ is the
Neumann Green function satisfying $-\mu_b^{-1} \nabla \cdot (\mu_b
\nabla g_N) = \delta({\bf x} - {\bf x}')$ with boundary condition
${\bf \hat n} \cdot \nabla g_N = 0$.

The initial surface screening current appears in the transverse
magnetic boundary condition \cite{Jackson,W03}: ${\bf K}({\bf r})
= (c/4\pi) {\bf \hat n} \times [{\bf H}_b({\bf r}) - {\bf
H}_c({\bf r})]$.  As indicated in Fig.\ \ref{fig1}, ${\bf H}_c$ is
the same as that prior to pulse termination, while ${\bf H}_b$ is
given by (\ref{3}):
\begin{equation}
{\bf K} = -\frac{c}{4\pi} {\bf \hat n}
\times (\nabla \Phi - {\bf H}_c)_{t=0},
\label{4}
\end{equation}
which, via (\ref{3}), determines ${\bf K}$ entirely in terms of
${\bf H}_c$.

To investigate the evolution of ${\bf K}$ for $t > 0$, we take
advantage of the rapid variation of the fields near the surface
with the local vertical coordinate $z$.  Thus, $z$-derivatives
dominate (\ref{2}), and to leading order in the small parameter
$\epsilon(t) = \sqrt{D_c t/L_c^2}$ \cite{foot1}, one obtains
\begin{equation}
-\partial_Z^2 {\bf A}^\perp + \partial_t {\bf A} = 0,\ z < 0,
\label{5}
\end{equation}
with initial condition ${\bf E}(0) = -(1/c) \partial_t {\bf A}(0)
= \sigma_c^{-1} {\bf K} \delta(z)$.  Here $Z = z/\sqrt{D_c}$,
${\bf A}^\perp = {\bf A} - {\bf \hat n} ({\bf \hat n} \cdot {\bf
A})$ is the tangential part of ${\bf A}$, and $\mu,\sigma,D$ are
treated as constants on either side of the boundary.  One may
choose $\Delta {\bf A} = 0$ for $t = 0$, and it follows
immediately that ${\bf \hat n} \cdot \Delta {\bf A} \equiv 0$:
$\Delta {\bf A}$ is purely transverse.

One requires now the boundary condition for (\ref{5}) at $Z =
0^-$.  For $t > 0$ let ${\bf A} = {\bf A}(0) + \Delta {\bf A}(t)$
and $\Phi = \Phi(0) + \Delta \Phi(t)$, where one treats $|\Delta
{\bf A}|/|{\bf A}(0)|, |\Delta \Phi|/|\Phi(0)| = {\cal
O}(\epsilon)$ and $|\partial_Z \Delta {\bf A}|/|\Delta {\bf A}| =
{\cal O}(1/\epsilon)$.  Keeping only leading terms \cite{foot1},
continuity of ${\bf \hat n} \times {\bf H}$ (the surface current
sheet now has finite thickness) and (\ref{4}) imply that
\begin{equation}
\sqrt{\frac{4\pi \mu_c}{\sigma_c}} {\bf K}
= -\partial_Z \Delta {\bf A}^\perp|_{Z=0^-}
+ \mu_c \sqrt{D_c} {\bf \hat n}
\times \nabla \Delta \Phi|_{Z=0^+},
\label{6}
\end{equation}
in which $\Delta \Phi$ is given by $\Delta {\bf A}(Z=0^-)$ via
(\ref{3}). In fact, estimating $\partial_Z = {\cal
O}[D_c^{1/2}/\epsilon(t) L_c]$, $g_N = {\cal O}(1/L_c)$, and ${\bf
\hat n} \times \nabla = {\cal O}(1/L_c)$, the ratio of the second
term on the right hand side to the first is ${\cal O}[\epsilon(t)
\mu_c/\mu_b]$. Their relative order is therefore
\emph{time-dependent} \cite{foot2}. In particular, at early-early
time $\epsilon(t) \ll \mu_b/\mu_c$ (i.e., $t \ll \tau_{\rm mag}$)
one may drop the second term. Lack of time dependence in ${\bf K}$
then leads to the simple homogeneous Neumann boundary condition
$\partial_z {\bf E} = 0$ for the electric field.  This is the
limit in which the early time analysis in Ref.\ \cite{W03} was
carried out and the $t^{-1/2}$ behavior of $V(t)$ derived.

Thus, for $\mu_c/\mu_b = {\cal O}(1)$ one gains nothing by
including the $\Delta \Phi$ term:  it is of the same order as
other ${\cal O}[\epsilon(t)]$ terms previously dropped from
(\ref{6}) \cite{foot1}.  However, if $\mu_c/\mu_b \gg 1$ this term
dominates for $\tau_{\rm mag} < t < \tau_e$, which now comprises a
large fraction of the early time regime. The remainder of this
paper is concerned with extending the theory into this regime.

With the $\Delta \Phi$ term, (\ref{6}) is nonlocal, coupling
$\Delta {\bf A}(Z=0^-)$ over the surface. In order to decouple
(\ref{6}) we \emph{diagonalize} it by seeking transverse vector
eigenfunctions ${\bm \alpha}_n$, $n=1,2,3,\ldots$, satisfying
\begin{equation}
\kappa_n {\bm \alpha}_n({\bf r})
= -\mu_c \sqrt{D_c} {\bf \hat n}
\times \nabla \psi_n({\bf r})
\label{7}
\end{equation}
where $\kappa_n$ is the corresponding eigenvalue, and ${\bf r}$ is
the surface coordinate.  The scalar $\psi_n$ is defined by
\begin{equation}
\psi_n = \hat {\cal L}^N_g \mu_b^{-1} {\bf \hat n}
\cdot \nabla \times {\bm \alpha}_n,
\label{8}
\end{equation}
where, to condense the notation, we define the Neumann Green
function operator via $\hat {\cal L}^N_g \phi({\bf r}) =
\int_{\partial V_c} d^2r' g_N({\bf r},{\bf r}') \phi({\bf r}')$.
Substituting (\ref{7}) into (\ref{8}), one obtains a \emph{scalar}
eigenvalue problem,
\begin{equation}
\kappa_n \psi_n = \hat {\cal L} \psi_n,\ \
\hat {\cal L} \equiv -\hat {\cal L}^N_g \mu_b^{-1}
\hat {\cal L}_\Delta,
\label{9}
\end{equation}
where the generalized surface Laplacian is $\hat {\cal L}_\Delta
\phi({\bf r}) = {\bf \hat n} \cdot \nabla \times \left[\mu_c
\sqrt{D_c} {\bf \hat n} \times \nabla \phi({\bf r}) \right]$. On a
sphere one obtains $\hat {\cal L}_\Delta = -\mu_c \sqrt{D_c}
L_c^{-2} {\bf L}^2$, where $L_c$ is now the radius, and ${\bf L} =
-i {\bf x} \times \nabla$ is the angular momentum operator
\cite{Jackson}.

As the ${\bm \alpha}_n$ are derived from the scalar $\psi_n$ via
(\ref{7}), they do not form a complete set of 2D vector fields.
The missing fields consist of the kernel of (\ref{7}) (the space
of functions $\kappa_n = 0$), spanned by all fields ${\bm \beta}$
with vanishing normal magnetic field, ${\bf \hat n} \cdot \nabla
\times {\bm \beta} = 0$.  It follows that ${\bm \beta} =
-\nabla^\perp \phi \equiv {\bf \hat n} \times {\bf \hat n} \times
\nabla \phi$ for some other scalar $\phi$. Let $\phi_n$, ${\bm
\beta}_n$ be chosen as eigenfunctions of the generalized
transverse Laplace equation,
\begin{eqnarray}
\lambda_n \phi_n({\bf r}) &=& -{\bf \hat n} \cdot \nabla \times
\left[\mu_c^{-1} D_c^{-1/2} {\bf \hat n} \times
\nabla \phi_n({\bf r}) \right]
\nonumber \\
{\bm \beta}_n({\bf r}) &=& -\nabla^\perp \phi_n,
\label{10}
\end{eqnarray}
with a new set of eigenvalues $\lambda_n$.

The set $\{{\bm \alpha}_n,{\bm \beta}_n \}$ forms a complete
basis, and we perform the \emph{magnetic surface mode expansion}
\begin{eqnarray}
\Delta {\bf A}({\bf x},t) &=& \sum_{n=1}^\infty
\left[A^{(1)}_n(Z,t) {\bm \alpha}({\bf r})
+ A^{(2)}_n(Z,t) {\bm \beta}({\bf r}) \right]
\nonumber \\
\sqrt{\frac{4\pi \mu_c}{\sigma_c}} {\bf K}({\bf r})
&=& \sum_{n=1}^\infty
\left[K^{(1)}_n {\bm \alpha}({\bf r})
+ K^{(2)}_n {\bm \beta}({\bf r}) \right].
\label{11}
\end{eqnarray}
From (\ref{7}), $\hat {\cal L}_g^N \mu_b^{-1} {\bf \hat n} \cdot
\nabla \times \Delta {\bm A} = \sum_n A^{(1)}_n \psi_n$, and from
(\ref{10}), ${\bf \hat n} \cdot \nabla \times (\mu_c^{-1}
D_c^{-1/2} \Delta {\bf A}) = \sum_n \lambda_n A^{(2)}_n \phi_n$
(and similarly for ${\bf K}$), hence appropriate orthogonality
relations for $\psi_n,\phi_n$ may be used to determine
$A_n^{(i)},K_n^{(i)}$.

Substituting (\ref{11}) into (\ref{5}) and (\ref{6}), complete
separation of the surface modes is achieved:
\begin{equation}
(\partial_t - \partial_Z^2) A^{(i)}_n = 0,\ Z < 0,
\label{12}
\end{equation}
with initial and boundary conditions
\begin{eqnarray}
-\partial_t A^{(i)}_n|_{t=0^+} &=& K^{(i)}_n \delta(Z)
\nonumber \\
(\partial_Z + \kappa_n \delta_{i1}) A^{(i)}_n|_{Z=0^-}
&=& -K^{(i)}_n.
\label{13}
\end{eqnarray}
The solutions are $A^{(i)}_n(Z,t) = -K^{(i)}_n H(Z,t;\kappa_n
\delta_{i1})$, where
\begin{eqnarray}
H(Z,t;\kappa) &=& \textstyle \frac{1}{\kappa}
\left[{\rm erfc}\left(\frac{|Z|}{\sqrt{4t}} \right)
- e^{\kappa^2 t - \kappa Z}
{\rm erfc}\left(\frac{2\kappa t - Z}{\sqrt{4t}} \right) \right]
\nonumber \\
H(Z,t;0) &=& \textstyle \sqrt{\frac{4t}{\pi}} e^{-Z^2/4t}
- |Z| {\rm erfc}\left(\frac{|Z|}{\sqrt{4t}} \right),
\label{14}
\end{eqnarray}
where ${\rm erfc}(x)$ is the complementary error function.

\begin{figure*}[tbh]

%use the following for two-column format
\includegraphics[scale=0.32]{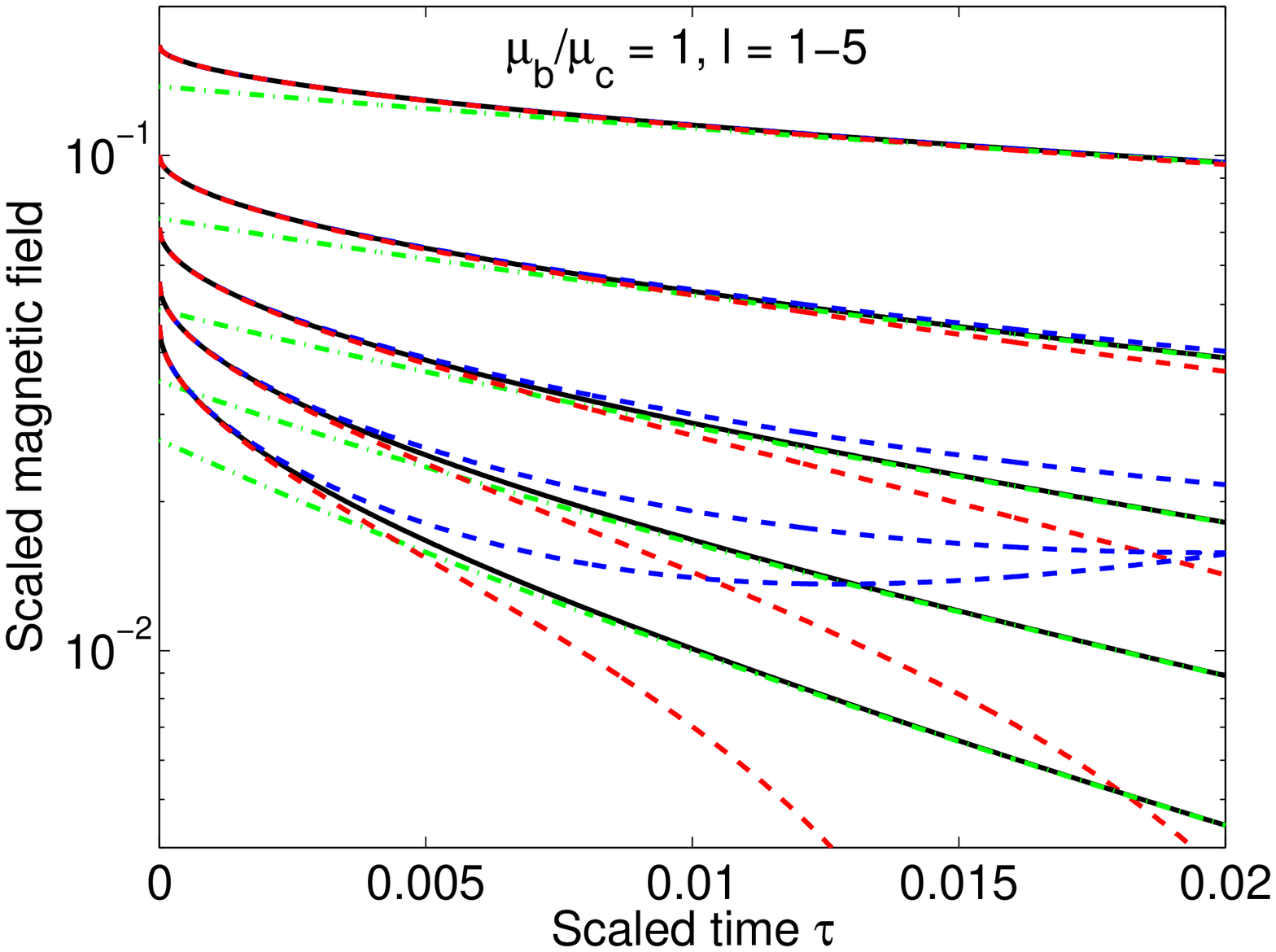}
\includegraphics[scale=0.32]{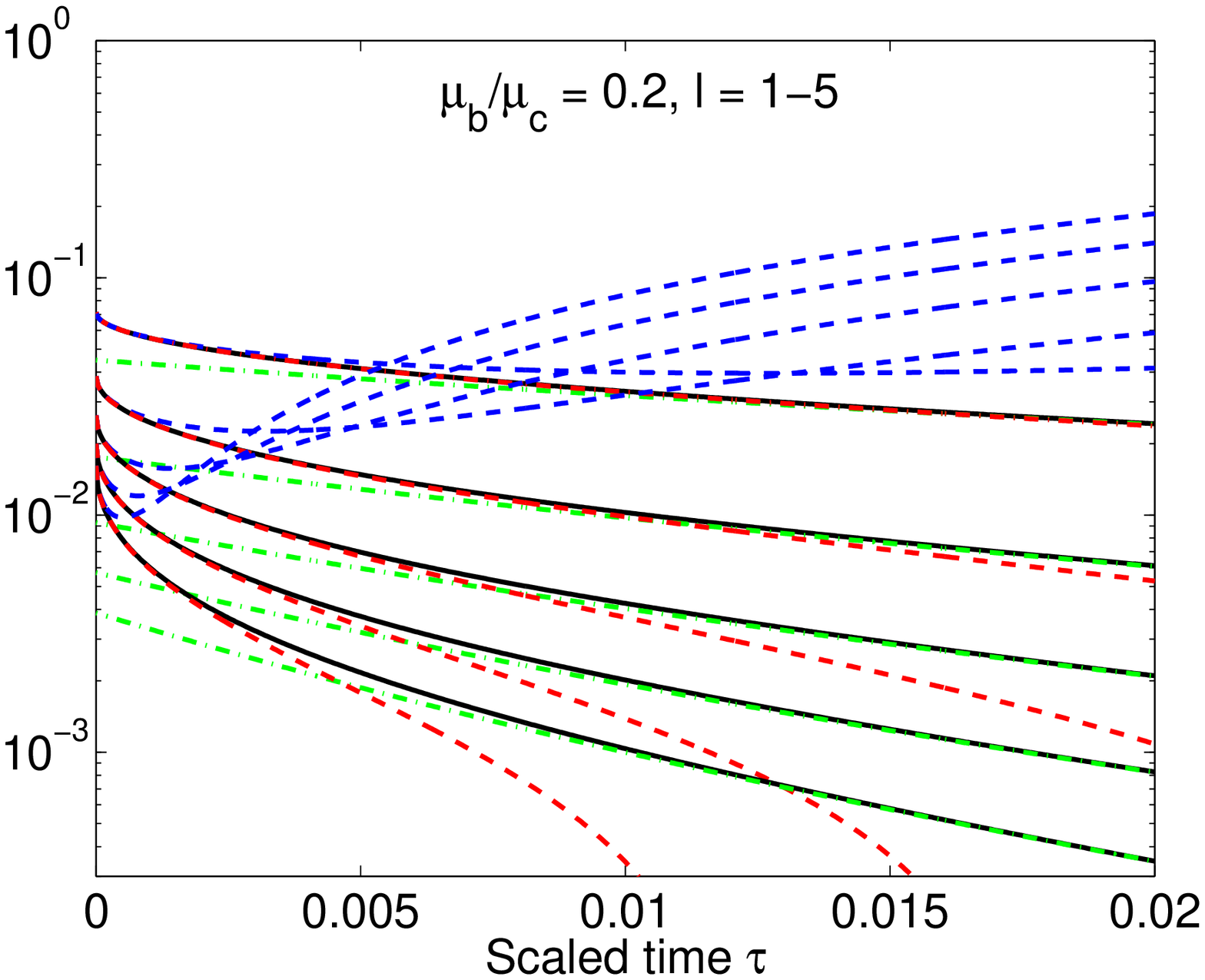}
\includegraphics[scale=0.32]{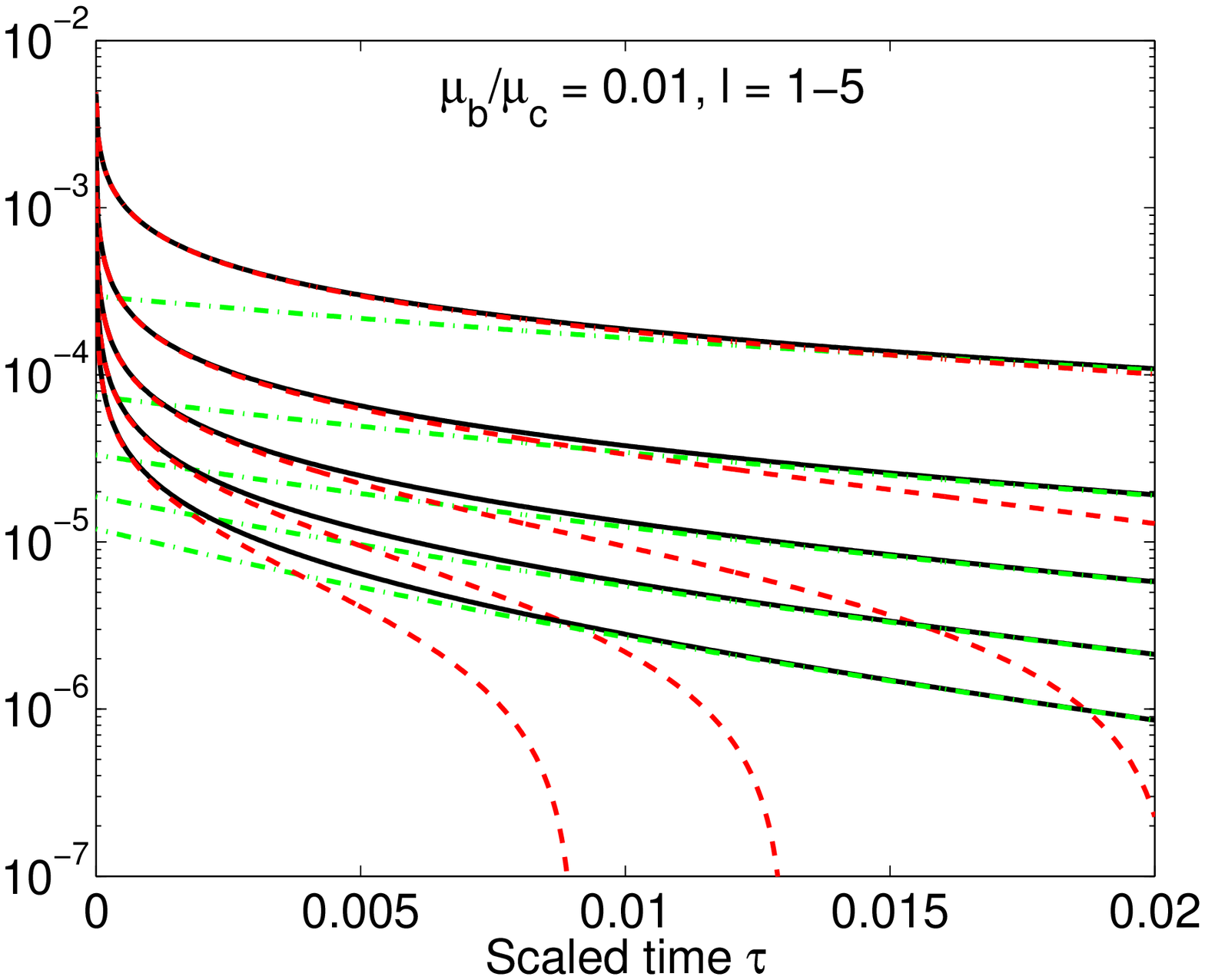}

%use the following for preprint format
%\includegraphics[scale=0.29]{prlfig3a.eps}
%\includegraphics[scale=0.29]{prlfig3b.eps}
%\includegraphics[scale=0.29]{prlfig3c.eps}

\caption{(COLOR) Comparison between exact and early time results
for spheres in a homogeneous background. Plotted is $H_l(\tau
\equiv t/\tau_c)$ for $1 \leq l \leq 5$ (curves for higher values
of $l$ lie lower on the plots). Left: $\mu_c/\mu_b = 1$.  Blue
(dashed) lines show the asymptotic early-early-time form
(\ref{19}); red (dashed) lines show the full early time solution;
black (solid) lines show exact results; and green (dash-dotted)
lines show (\ref{21}) truncated at the first 3 terms (an
approximation that might emerge from a late time perturbative
approach \protect\cite{WL03}). The two early time curves, though
differing in detail, exhibit roughly the same accuracy, with
interval of validity shrinking as $L_{lm}/L_c = 1/l^2$, as
predicted. In realistic applications, the transmitted field in the
target region will be fairly uniform, and one may expect $l = 1$
to dominate, with small corrections from higher $l$. The union of
early and late time approximations then provide a rather accurate
description of the full signal. Center: $\mu_c/\mu_b = 5$.  The
full early time form now clearly exhibits greatly extended
accuracy over the early-early time power law. Right: $\mu_c/\mu_b
= 100$. The full early-time form is indistinguishable from the
late-early time form (\ref{19}), the early-early time interval
being invisibly narrow.}

\label{fig3}
\end{figure*}

Fields \emph{external} to the target are obtained by extending
${\bm \alpha}_n,{\bm \beta}_n$ into the exterior space. First, let
\begin{eqnarray}
\psi_n({\bf x}) &=& -\frac{1}{\kappa_n} \int_{\partial V_c}
d^2r' g_N({\bf x},{\bf r}') \frac{1}{\mu_b}
\hat {\cal L}_\Delta \psi_n({\bf r}')
\nonumber \\
\phi_n({\bf x}) &=& \int_{\partial V_c} d^2r'
g_D({\bf x},{\bf r}') \phi_n({\bf r}'),
\label{15}
\end{eqnarray}
in which $g_D({\bf x},{\bf x}')$ is the Dirichlet Green function
satisfying $-\sigma_b^{-1} \nabla \cdot (\sigma_b \nabla g_D) =
\delta({\bf x}-{\bf x}')$ and vanishing on $\partial V_c$, and
${\bf x}$ is no longer restricted to the surface.  These
definitions guaranteed continuity at the boundary. The vector
eigenfunctions are now extended by solving
\begin{eqnarray}
\mu_b \nabla \times {\bm \alpha}_n({\bf x})
= -\nabla \psi_n({\bf x}),\
\nabla \cdot (\sigma_b {\bm \alpha}_n) = 0,
\label{16}
\end{eqnarray}
while imposing continuity of ${\bm \alpha}^\perp_n$ at the
surface, and
\begin{equation}
{\bm \beta}_n({\bf x}) = -\nabla \phi_n({\bf x}).
\label{17}
\end{equation}
Since $\nabla \times {\bm \beta}_n = 0$ it does not contribute to
the external magnetic field, and hence to any inductive
measurement.  The external field is now simply
\begin{equation}
\Delta {\bf A}({\bf x},t) = \sum_{n=1}^\infty
\left[A^{(1)}_n(0^-,t) {\bm \alpha}_n({\bf x})
+ A^{(2)}_n(0^-,t) {\bm \beta}_n({\bf x}) \right],
\label{18}
\end{equation}
with time-dependence given by the internal field on the boundary,
hence governed by the function
\begin{eqnarray}
&&H(0,t;\kappa) = \frac{1}{\kappa} \left[1 - e^{\kappa^2 t}
{\rm erfc}\left(\kappa \sqrt{t} \right) \right]
\label{19} \\
&&\to \begin{cases}
\sqrt{\frac{4t}{\pi}} \left[1 - \frac{1}{2}
(\pi \kappa^2 t)^{1/2} + {\cal O}(\kappa^2 t) \right],
& \kappa^2 t \ll 1 \\
\frac{1}{\kappa} \left\{1 - (\pi \kappa^2 t)^{-1/2}
+ {\cal O}[(\kappa^2 t)^{-3/2}] \right\} & \kappa^2 t \gg 1.
\end{cases}
\nonumber
\end{eqnarray}
The time derivative, entering the electric field and voltage,
displays the promised $t^{-1/2}$ and $t^{-3/2}$ power laws in the
two limits.  If $\psi_n$ varies on scale $L_n \leq L_c$, one may
estimate from (\ref{9}) $\kappa_n = {\cal O}[\tau_{\rm mag}^{-1/2}
(L_c/L_n)]$, hence crossover point $\tau_{{\rm mag},n} \equiv
1/\kappa_n^2 = {\cal O}[\tau_{\rm mag} (L_n/L_c)^2]$. One expects
$L_n = L_c$ only for the fundamental mode, hence (\ref{1})
actually represents an \emph{upper bound} on the spectrum of
crossover times $\tau_{{\rm mag},n}$.

The exact solution for a homogeneous sphere in a homogeneous
background serves to clarify all of the above concepts.  The
operators $\hat {\cal L}_g^N$ and $\hat {\cal L}_\Delta$ now
commute and may be simultaneously diagonalized using spherical
harmonics: $\hat {\cal L}_g^N Y_{lm} = Y_{lm}/(l+1)L_c$, $\hat
{\cal L}_\Delta Y_{lm} = -[\mu_c \sqrt{D_c} l(l+1)/L_c^2] Y_{lm}$.
Thus, using $\psi_{lm} = \phi_{lm} = L_c^{-2}Y_{lm}$ one obtains
$\kappa_{lm} = l \tau_{\rm mag}^{-1/2}$ (hence mode length scale
$L_{lm} = L_c/l$), and $\lambda_{lm} = l(l+1)/L_c^2\mu_c
\sqrt{D_c}$. The vector functions are ${\bm \alpha}_{lm}({\bf x})
= -i\mu_b \sqrt{(l+1)/l} (L_c^l/x^{l+1}) {\bf X}_{lm}$, ${\bm
\beta}_{lm}({\bf x}) = -i \sqrt{(l+1)/l} \nabla \times
[(L_c^l/x^{l+1}) {\bf X}_{lm}]$, where ${\bf X}_{lm} =
[l(l+1)]^{-1/2} {\bf \hat L} Y_{lm}$ are the vector harmonics
\cite{Jackson}. The exact solution associated with the magnetic
modes ${\bm \alpha}_{lm}$ is governed by the usual bulk
exponentially decaying mode expansion \cite{WL03,W03}, from which
one obtains $A_{lm}^{(1)}(t) = K^{(1)}_{lm} \sqrt{4\tau_c} [H_l(0)
- H_l(t/\tau_c)]$, where
\begin{equation}
H_l(\tau) = \sum_{n=1}^\infty \frac{j_l(\zeta_{ln})^2}
{j_l(\zeta_{ln})^2 - j_{l+1}(\zeta_{ln}) j_{l-1}(\zeta_{ln})}
\frac{e^{-\zeta_{ln}^2 \tau}}{\zeta_{ln}^2},
\label{20}
\end{equation}
where $j_l(x)$ are the spherical Bessel functions, and the scaled
decay rates are given by the roots of
\begin{equation}
0 = \frac{\mu_b}{\mu_c} \zeta_{ln} j_{l-1}(\zeta_{ln})
+ l\left(1 - \frac{\mu_c}{\mu_b} \right) j_l(\zeta_{ln}).
\label{21}
\end{equation}
In Fig.\ \ref{fig3} $H_l(t/\tau_c)$ is compared to the early time
prediction $H_l(t/\tau_c) \approx H_l(0) - (4\tau_c)^{-1/2}
H(0^-,t;\kappa_{lm})$, with $H_l(0) = (\mu_b/2\mu_c)/[l +
(l+1)\mu_b/\mu_c]$, for various $l$, $\mu_c/\mu_b$.

The author is indebted to E. M. Lavely for numerous discussions.
The support of SERDP, through contract No.\ DACA 72-02-C-0029, is
gratefully acknowledged.

\end{document}